\documentclass[biblatex,english,runningheads]{lniArXiv}
\usepackage{tikz}
\usetikzlibrary{shapes,arrows.meta,decorations.text,calc,positioning}

\begin{document}
\title[Agricultural Industry Initiatives on Autonomy]{Agricultural Industry Initiatives on Autonomy}
\subtitle{How collaborative initiatives of VDMA and AEF can facilitate complexity in domain crossing harmonization needs.} % if needed
 \author[1]{Georg Happich}{georg.happich@hs-kempten.de}{0009-0001-4975-3703}
 \author[2]{Alexander Grever}{alexander.grever@krone.de}{0009-0006-0694-1145}
 \author[3]{Julius Schöning}{j.schoening@hs-osnabrueck.de}{0000-0003-4921-5179}
 \affil[1]{Kempten University of Applied Sciences\\Faculty of Mechanical Engineering\\Bahnhofstraße 61\\DE-87435 Kempten\\Germany}
 \affil[2]{Bernard Krone GmbH \& Co. KG \\Heinrich-Krone-Straße 10\\DE-48480 Spelle\\Germany}
 \affil[3]{Osnabrück University of Applied Sciences\\Faculty of Engineering and Computer Science\\DE-Albrechtstr. 30\\Germany}
\maketitle

\begin{abstract}
The agricultural industry is undergoing a significant transformation with the increasing adoption of autonomous technologies. Addressing complex challenges related to safety and security, components and validation procedures, and liability distribution is essential to facilitate the adoption of autonomous technologies. This paper explores the collaborative groups and initiatives undertaken to address these challenges. These groups investigate inter alia three focal topics: 1) describe the functional architecture of the operational range, 2) define the work context, i.e., the realistic scenarios that emerge in various agricultural applications, and 3) the static and dynamic detection cases that need to be detected by sensor sets. Linked by the \textit{Agricultural Operational Design Domain} (Agri-ODD), use case descriptions, risk analysis, and questions of liability can be handled. By providing an overview of these collaborative initiatives, this paper aims to highlight the joint development of autonomous agricultural systems that enhance the overall efficiency of farming operations.
\end{abstract}
\begin{keywords}
Autonomy \and  Agriculture \and  Safety \and  Security \and  Liability \and Development 
\end{keywords}
%%% Beginn des Artikeltexts
\section{Introduction and Objective}
For over a decade, autonomous solutions were mainly presented by small and medium-sized companies, by innovative start-ups, and as the result of research projects. At the same time, the so-called traditional big players in the agricultural industry acted reserved. However, projects for inspecting technology solutions and understanding legal opportunities and constraints were presented, e.g., by AGCO, John Deere, and CNHi. During the past two years, this strategy has been drastically changed \cite{Hindman2025}. Following, but not limited to, the success of the European industry collaborations \textit{Advanced Automation and Autonomy} (A3) by CLAAS, AgXeed, and Amazone and \textit{Combined Powers} by Lemken and Krone, there is a remarkable push towards autonomous field operations. Thus, autonomous solutions have become a significant topic in the agricultural industry.

\pagebreak
Most of the constraints for autonomous operations in agriculture can be traced back to three questions, which contain the most important answers for autonomy in agriculture:
\begin{itemize}
\item How must the role and requirements for functional safety in autonomous systems, characterized by the absence of a local operator, the driver, be described?
\item What components and setup of sensor sets allow for autonomous operations, and how can sensor sets be evaluated for various purposes in the agricultural industry?
\item How can autonomous operation, responsibility, and the underlying liability aspects be distributed over operational agricultural systems, especially in classic tractor-implement combinations?
\end{itemize}

Several agricultural industry associations have launched specific projects to answer the relevant aspects of these questions. For example, the \textit{European Agricultural Machinery Association} (CEMA) is running the \textit{Safety in Autonomous Functions initiative} called PT4, and the \textit{Agricultural Industry Electronics Foundation} (AEF e.V.) started the project team \textit{Autonomy in Agriculture} (AEF AUT). Other initiatives concentrate on obviously contributing aspects, such as the security of services, broad bandwidth wireless and long-distance communication, and the ethical impacts of autonomous operations. In an informative matter, this short paper will report and summarize these activities.

\section{Collaborative Initiatives and Working Groups}
In early 2024, the \textit{Technical Committee on Electronics} (TAE) of the \textit{German Machinery and Equipment Manufacturers Association} (VDMA), in collaboration with the AEF, hosted a workshop aimed at discussing and highlighting the role of autonomy in agriculture, specifically focusing on highly automated agricultural machinery. This workshop addressed the expanding importance of autonomous machinery and the regulatory and practical requirements for bringing such machinery to market, assigning topics to the AEF and VDMA committees for further exploration, especially considering the already existing initiative PT4 by CEMA.

The workshop also examined the limitations of current testing frameworks, such as ISO 18497, which outlines test cases adopted from other initiatives, like the INRAE \textit{Agricultural Robotic Performance Assessment} (APRA) \cite{Vargas2023}, NARO \textit{Autonomous Agri-Machinery Test} \cite{NAFRO2021}. The findings indicated that the existing specifications in ISO 18497 are insufficient for the safe market entry of autonomous agricultural machinery. In response, the joint committee decided to establish a dedicated working group under the leadership of the technical committee on electronics, focusing on sensor testing with a particular focus on human detection. This working group has been tasked with developing enhanced guidelines for testing sensor systems to ensure effective object detection for personal safety.

\pagebreak
The group's specific objectives include:
\begin{itemize}
\item  Demonstrating the detection capabilities of sensors for object identification,
\item  Defining objects in a way that ensures personal safety in agricultural settings,
\item  Conducting tests on complete sensor systems, including sensors, software, and safety-related functions, and
\item  Achieving at least an ISO \textit{Technical Specification} (TS) level for documentation.
\end{itemize}

The working group comprises representatives from OEM, university faculty and researchers, VDMA representatives, sensor manufacturers, system suppliers, functional safety experts, and testing professionals specializing in machine testing.

In addition, the AUT project team focuses on challenges associated with interoperability, usability, and the fundamental removal of technical barriers to enable the use of existing standards, such as \textit{Tractor Implement Management} (TIM) in the context of autonomy. The alignment of the project is the result of a two-part workshop series initiated by the AEF in 2023. In this workshop series, the AEF steering committee discussed and developed the strategic direction and goals of the project together with its member companies, universities and research institutions, and representatives of the VDMA. The objectives of the AUT project include:
\begin{itemize}
\item  Definition of an interoperable autonomous architecture. A key objective is to create a framework that enables different autonomous systems to communicate with each other and work together seamlessly.
\item Identification and removal of technical barriers to autonomy and ensuring ease of use for customers together with existing AEF teams
\item Changing existing guidelines and standards: The project team will review current standards and guidelines and advocate for changes that enable autonomy and improve interoperability.
\item Creating new standards for autonomy: In cases where existing guidelines are insufficient, the AUT project team will develop new standards to fill the gaps needed for a holistic approach to autonomy.
\end{itemize}

The project team launched in April 2024 and is creating the first draft of an interoperable architecture, which will then be shared with all AEF members for review. 
One of the primary challenges for all groups and initiatives on this topic is the significant variability in environmental conditions, obstacles, and vegetation profiles across agricultural applications ranging from indoor livestock via yard to in-field systems. For instance, in fields where low or even no vegetation can be expected in typical soil cultivation, harvest processes mostly involve fundamentally different conditions. Similarly, in dry environments with straw or soil, detection systems for autonomous operation must contend with high levels of dust, whereas in humid conditions, these systems are more likely to encounter mud, water droplets, and fog. Consequently, it is essential to define and establish tools to broadly characterize the machine's system operating conditions while also narrowing these to the specific environmental conditions described.

In the automotive industry, the concepts of an \textit{Operational Design Domain} (ODD) and a \textit{Dynamic Driving Task} (DDT), according to SAE~J3016, are used to limit vehicle automation's complexity and achieve a feasible number of functional safety-related test cases \cite{Schoening2024,Schoening2023}. Briefly, the ODD categorizes and restricts the environment in which a vehicle operates, while the DDT outlines the operational tasks, i.e., lateral and longitudinal motion control required for vehicle control. The tactical function that triggers the operation task includes the object, event detection, and response.

When attempting to apply the ODD and DDT frameworks to the agricultural sector, it quickly becomes evident that this approach is relevant for all three working groups. For each of them, a framework like this is vital to abstract core questions like:
\begin{enumerate}
\item How can the operational range of the system be defined? --- function architecture 
\item  What realistic scenarios emerge in various agricultural applications? --- application context
\item  Which static and dynamic use cases occur and need to be tested? --- sensor testing
\end{enumerate}

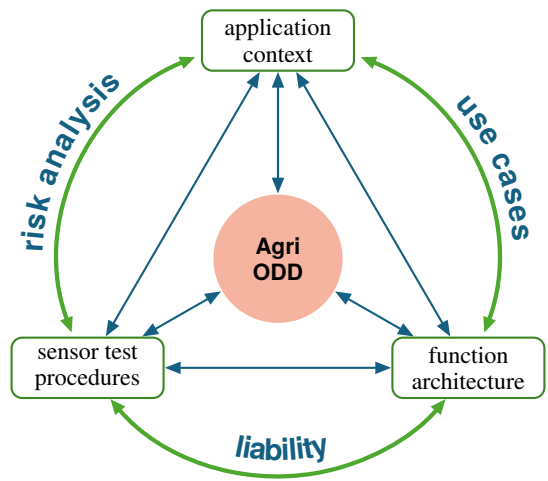
\begin{figure}[h]
  \centering
  \definecolor{greenA}{RGB}{78,167,46}
  \definecolor{blueA}{RGB}{21,96,130}
  \definecolor{redA}{RGB}{244,180,160}

  \begin{tikzpicture}[RecNode/.style={draw=greenA!80!black, thick, rectangle, rounded corners, minimum width = 2.0cm, minimum height = 0.8cm, inner sep=0pt, align=center}, CirNode/.style={fill=redA, thick, circle,  rounded corners, minimum width = 1.7cm, minimum height = 1.7cm, inner sep=0pt, align=center, font=\sffamily\bfseries}, NArrows/.style={draw=blueA, arrows = {Stealth[length=6pt, inset=0pt]-Stealth[length=6pt, inset=0pt]},thick}, CArrows/.style={draw=greenA, arrows = {Stealth[length=6pt, inset=0pt]-Stealth[length=6pt, inset=0pt]}, ultra thick,}]
    \def\myRadius{2.886}
    \def\myDiameter{\myRadius*2}

    \node[draw=none, circle, minimum width = \myDiameter cm, minimum height = \myDiameter cm, inner sep=0pt] (helpCirc){};
    \node[RecNode] at (helpCirc.90) (app_context) {application\\ context};
    \node[RecNode] at (helpCirc.-30) (sensor_test) {function \\ architecture};
    \node[RecNode] at (helpCirc.210) (func_arch) {sensor test \\ procedures};
    \node[CirNode] at (helpCirc.center) (odd) {Agri\\ ODD};
    
    \draw[NArrows] (odd) -- (app_context);
    \draw[NArrows] (odd) -- (sensor_test);
    \draw[NArrows] (odd) -- (func_arch);
    
    \draw[NArrows] (sensor_test) -- (app_context);
    \draw[NArrows] (func_arch) -- (app_context);
    \draw[NArrows] (func_arch) -- (sensor_test);

    \draw[CArrows, postaction={decorate,decoration={raise=1.5ex,text along path,text align=center,text={|\sffamily\bfseries\large\color{blueA}|  use cases}}}] ([shift=(68:\myRadius cm)]0,0) arc (68:-20:\myRadius cm);
    \draw[CArrows, postaction={decorate,decoration={raise=1.5ex,text along path,text align=center,text={|\sffamily\bfseries\large\color{blueA}|  risk analysis}}}] ([shift=(200:\myRadius cm)]0,0) arc (200:112:\myRadius cm);
    \draw[CArrows, postaction={decorate,decoration={raise=1.5ex,text along path,text align=center,text={|\sffamily\bfseries\large\color{blueA}|  liability}}}] ([shift=(-140:\myRadius cm)]0,0) arc (-140:-40:\myRadius cm);

\end{tikzpicture}
 \caption{Interplay of groups and initiatives that contribute to agricultural autonomy systems, all linked by the \textit{Agricultural Operational Design Domain} (Agri-ODD).}\label{fig:Interplay}
\end{figure}

Each question leads to one focal topic: functional architecture, work context, and sensor testing, illustrated in Figure~\ref{fig:Interplay}. In addition to these focal topics, there are indirect effects on the general interdependencies and challenges associated with autonomous agricultural machinery and its path to market. The center of all activities is the \textit{Agricultural ODD} (Agri-ODD), as a definition made by the manufacturer and machine, which can be instrumental in managing the high complexity of agricultural operations and their inherent interdependencies, making these challenges more tractable. The work of all the abovementioned groups and initiatives contribute to a focal topic that the Agri-ODD strongly links. Empowered by the Agri-ODD and the three focal topics, use case description, risk analysis, and questions of liability can be handled.

\section{Groups and Initiatives in Detail}
Focusing on sensor testing with a particular focus on human detection, a working group from the VDMA is developing enhanced guidelines for testing sensor systems to ensure effective object detection for personal safety. Within this working group, the top-down perspective, i.e., starting from the Agri-ODD to the sensor set, and the bottom-up perspective, i.e., starting from the standards, norms, and physical measurement principle to the sensor set, are considered.

From a bottom-up perspective, existing standards and test protocols from other industries are being scrutinized for their applicability in the agricultural sector. This scrutiny is very valuable when defining the requirements, which focus on sensor sets and systems' necessary functionality. Numerous standards and regulations are being revised to meet these upcoming requirements for autonomous solutions. For example, the adaptation of the Machinery Directive 2006/42/EC can be mentioned. While IEC 61496, as a harmonized standard for sensors, emphasizes mostly indoor and industrial use cases, IEC TS 62998 is a valuable input for agricultural applications due to its specific focus on outdoor scenarios. Additionally, system analysis in line with ISO 25119 and ISO 13849 is required, which makes this standard provide an intrinsic assessment of both hardware and software. Consequently, one key objective is to analyze it and identify and address gaps specific to agricultural applications. The immediate ongoing goal is to create a feature list defining essential measurement parameters, accuracy requirements, and test object specifications adapted for agrarian needs.

Starting from the Agri-ODD, the top-down perspective, the operational conditions for autonomous agricultural systems are systematically evaluated, particularly concerning object detection and the operational factors influencing this detection. The evaluation would benefit from a standardized Agri-ODD framework, illustrated as the center element in Figure~\ref{fig:Interplay}. However, since no standardized Agri-ODD framework exists, applying ODD's methodological approaches is still practical. Thus, a second ongoing objective of the working group is to create a detection-centered attribute list that maps environmental factors and characteristics like crop type, soil condition, and field topology to a few features for the sensor sets and systems. As work progresses, the variability of attributes and the increasing unpredictability as conditions approach real-world field applications pose a growing challenge to standardizing sensor testing, particularly regarding sensor systems. Based on a detection-centered attribute list, systematic testing methods like the parameter-based testing method \cite{Komesker2024}, along with methods from, e.g., electronics domain such as the design of experiments, orthogonal arrays or equivalence partitioning, and stochastic variation, will be feasible.

The central guiding question addressed by AUT is how to ensure the interoperability of autonomous agricultural systems through functional architectural standards and guidelines. Therefore, the project team developed a working method like classic systems engineering approaches during the first face-to-face workshop, which consists of the subsequent steps. The project team initially identifies and defines user roles and use cases within an autonomous system to create a common taxonomy and understanding as a basis for all members. In the second step, necessary architectural components and functional requirements are derived from use cases to obtain a notional autonomous architecture. At this point, each architectural component is defined by a description of its functionality and the required input and output information. Subsequently, the functional requirements of the software components are refined, and an initial interface definition is derived. After that, the use cases are traced through the architecture to validate the notional architecture. If the project team finds inconsistencies, the iteration is repeated, deriving the necessary software components and their specifications. After successful validation, the project team determines which interfaces between architecture components should be public or company-proprietary. Based on this, concrete safety, security, and functional requirements are specified for the public interfaces, and guidelines and standards are adapted together with existing AEF teams to fulfil the requirements for autonomy. A gap analysis is further carried out to determine any necessary new standard for interoperability.

\section{Outlook and Further Research}
The development of autonomous agricultural systems is gaining momentum, with various initiatives and working groups contributing. A standardized Agri-ODD framework should be defined as a crucial framework for managing the complexity of agrarian operations. Further research is needed to address the significant variability in environmental conditions, obstacles, and vegetation profiles. Developing detection-centered attribute lists and systematic testing methods will be essential for ensuring effective object detection and personal safety. Additionally, the creation of architectural standards and guidelines for interoperability will be critical for providing seamless communication and collaboration between different autonomous systems. The VDMA will extend the activity to the ISO level. Future joint research should focus on creating collaborative risk assessments and a responsibility map for questions of liability for autonomous agriculture systems.

\printbibliography

\end{document}